\documentclass[twocolumn,showpacs,preprintnumbers,amsmath,amssymb,prb]{revtex4-1}
\usepackage{amsfonts}
\usepackage[english]{babel}
\usepackage[T1]{fontenc}
\usepackage{times}
\usepackage{mathrsfs}
\usepackage{graphicx}
\usepackage{dcolumn}
\usepackage{bm}
\usepackage[colorlinks,bookmarks=true,citecolor=blue,linkcolor=red,urlcolor=blue]{hyperref}
\usepackage[tight, FIGTOPCAP, hang, raggedright, nooneline]{subfigure}

\newcommand\ket [1] {|#1 \rangle }
\newcommand\bra [1] {\langle #1 |}

\newcommand{\av}[1]{\langle #1\rangle}
\newcommand{\bb}[1]{\mathbf{#1}}

\begin{document}

\title{Flat bands with higher Chern number in pyrochlore slabs}

\author{Maximilian Trescher}
\author{Emil J. Bergholtz}\email{Author to whom correspondence should be addressed: ejb@physik.fu-berlin.de}
\affiliation{Dahlem Center for Complex Quantum Systems and Institut f\"ur Theoretische Physik, Freie Universit\"at Berlin, Arnimallee 14, 14195 Berlin, Germany}\
\date{\today}

\begin{abstract}
A large number of recent works point to the emergence of intriguing analogs of fractional quantum Hall states in lattice models due to effective interactions in nearly flat bands with Chern number $C=1$. Here, we provide an intuitive and efficient construction of almost dispersionless bands with higher Chern numbers. Inspired by the physics of quantum Hall multilayers and pyrochlore-based transition-metal oxides, we study a tight-binding model describing spin-orbit coupled electrons in $N$ parallel kagome layers connected by apical sites forming $N-1$ intermediate triangular layers (as in the pyrochlore lattice). For each $N$, we find finite regions in parameter space giving a virtually flat band with $C=N$. We analytically express the states within these topological bands in terms of single-layer states and thereby explicitly demonstrate that the $C=N$ wave functions have an appealing structure in which layer index and translations in reciprocal space are intricately coupled. This provides a promising arena for new collective states of matter. \end{abstract}

\maketitle

{\it Introduction.} Topological insulators, predicted theoretically\cite{topo} a few years 
ago and subsequently experimentally observed,\cite{topoexp} have attracted enormous amounts of interest. As these systems can usually be understood in a single-particle picture, an intriguing question is whether interactions can lead to qualitative new phenomena. Evidence that this is indeed the case has been accumulating during the past year. In a key step it was shown that {\it e.g.,} an appropriate combination of ferromagnetism and spin-orbit coupling can lead to nearly flat
bands with unit Chern number, $C=1$, in itinerant lattice systems.\cite{chernins1,chernins2,chernins3} Given the flat bands, these systems are likely to host interesting strongly correlated states and, at least in theory, this opens up a number of intriguing perspectives including high-temperature fractional quantum Hall states\cite{chernins1}. Indeed, numerical exact diagonalization studies convincingly show the existence of such states.\cite{cherninsnum1,cherninsnum2,bosons,nonab1,nonab2,nonab3,andreas} While the list of flat band models with $C=1$ is still growing\cite{c1a,c1b,c1c,c1d,c1e} and a better understanding of the relation between these bands and Landau levels\cite{andreas,qi,bands,cherncf,nonab1,cherntt} is developing, bands with higher Chern number could host qualitatively new phases of matter\cite{cn} as they have no direct analogue in the continuum.\cite{multilayers} In this context, two very recent papers made interesting progress by introducing models harboring relatively flat bands with $C=2$.\cite{c2,c2num}

At the same time there is a tremendous experimental development and theoretical interest in the physics of transition metal oxides, especially the iridates, where the effect of spin-orbit coupling is profound.\cite{hk,irtopo} Among these materials there are examples such as A$_2$Ir$_2$O$_7$ (A is a rare-earth element) where the relevant effective low-energy degrees of freedom are conducting 5d electrons on the Ir$^{4+}$ ions that live on the geometrically frustrated pyrochlore lattice formed by corner-sharing tetrahedra.\cite{pesin,weyl,fiete,kim} Interestingly, an anomalous Hall effect has been observed in the metallic pyrochlore.\cite{pyroqh}

\begin{figure}
\includegraphics[width=0.97\linewidth]{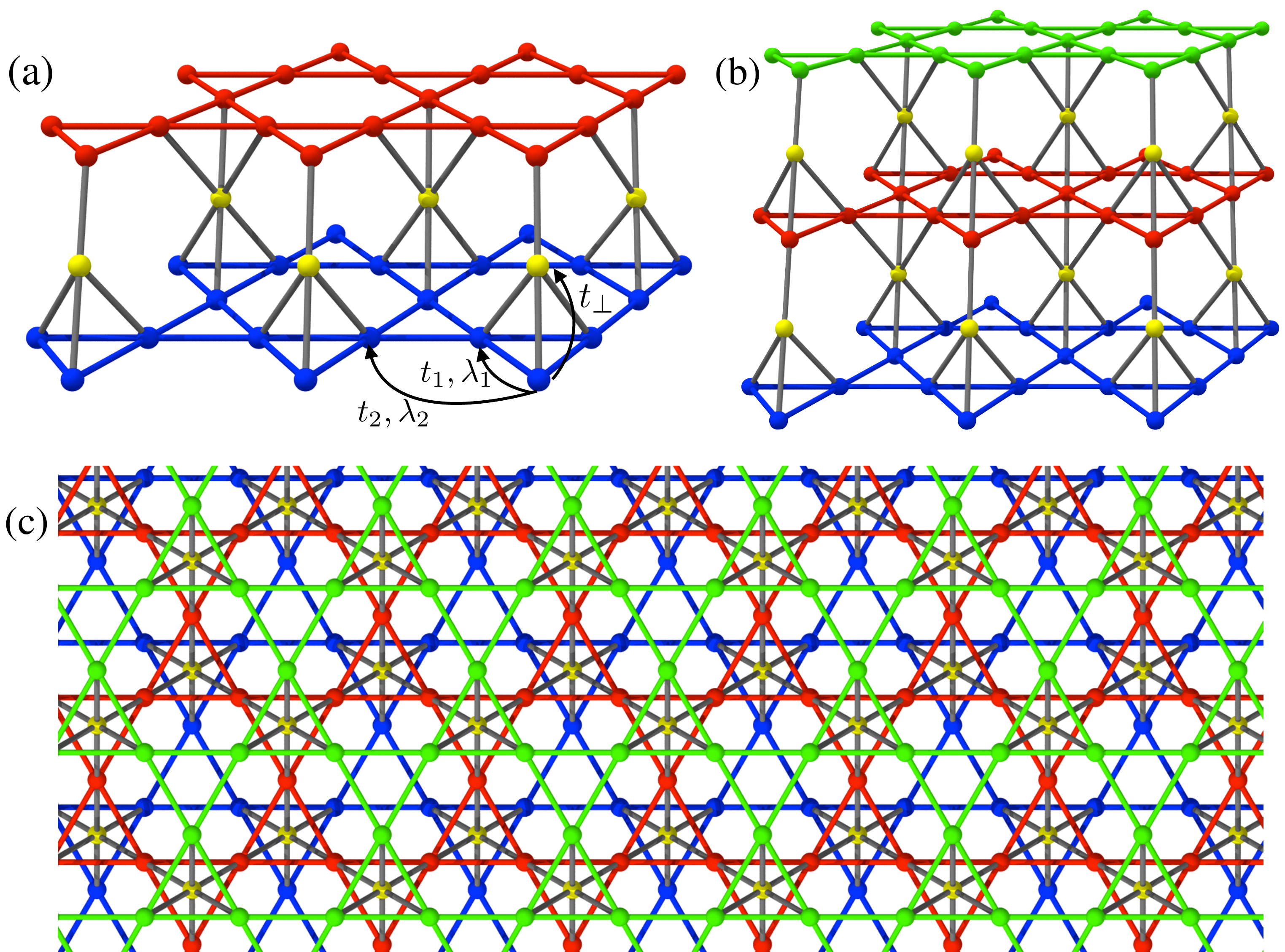}
\caption{(Color online) The kagome bilayer (a) and trilayer (b). The layers are colored differently for clarity and the considered hopping processes are indicated by arrows in (a). A 2D projection of the trilayer is shown in (c) and illustrates the general structure of the pyrochlore lattice: the projection has a three-fold periodicity (kagome layer $m+3n$ has the same projection as layer $m$).} 
\label{fig:lattice}
\end{figure} 

\begin{figure*}[t!]
    \subfigure[~]{\includegraphics[width=0.33\textwidth]{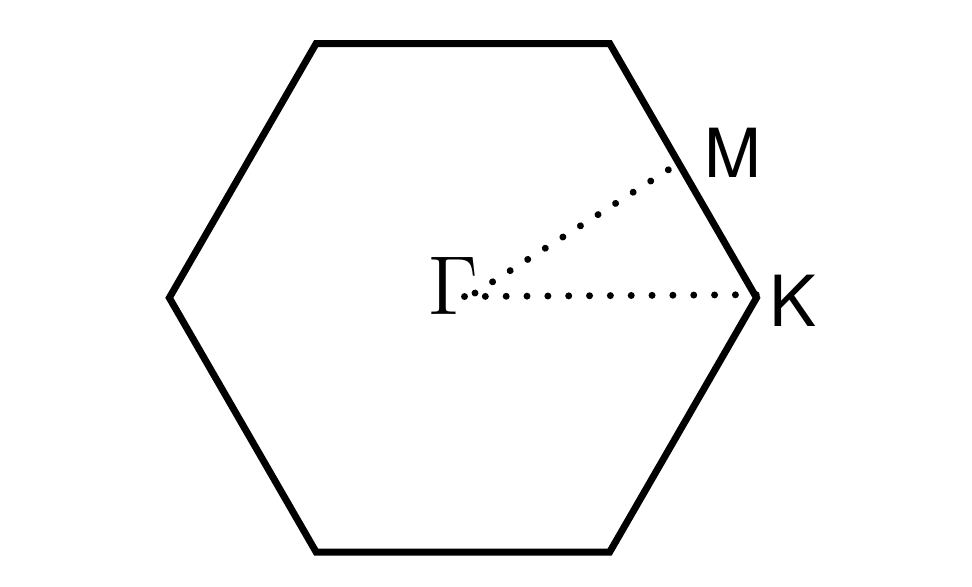}}
    \subfigure[~]{\includegraphics[width=0.33\textwidth]{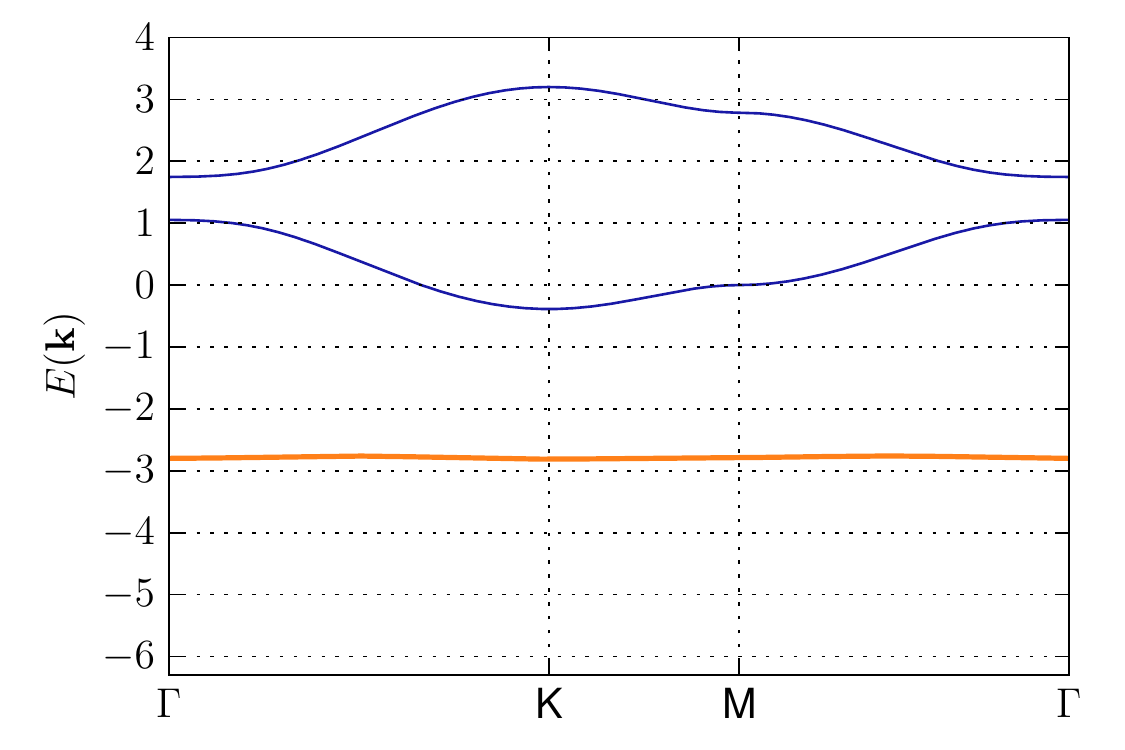}}
    \subfigure[~]{\includegraphics[width=0.33\textwidth]{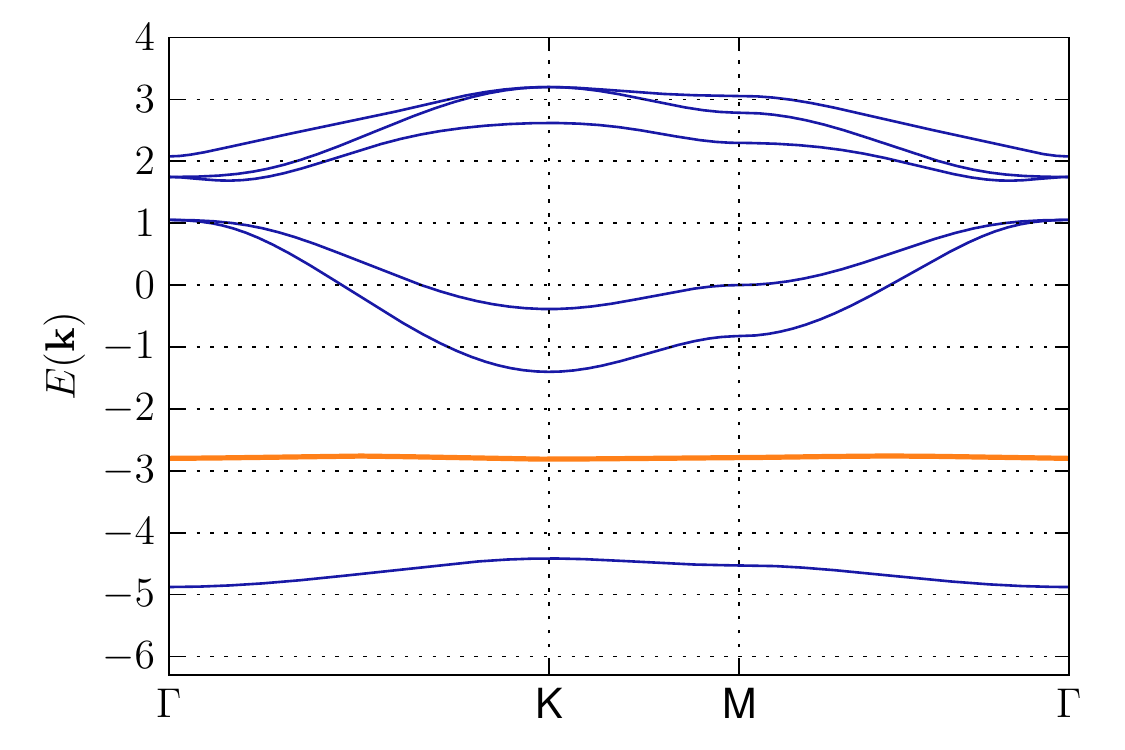}}
    \subfigure[~]{\includegraphics[width=0.33\textwidth]{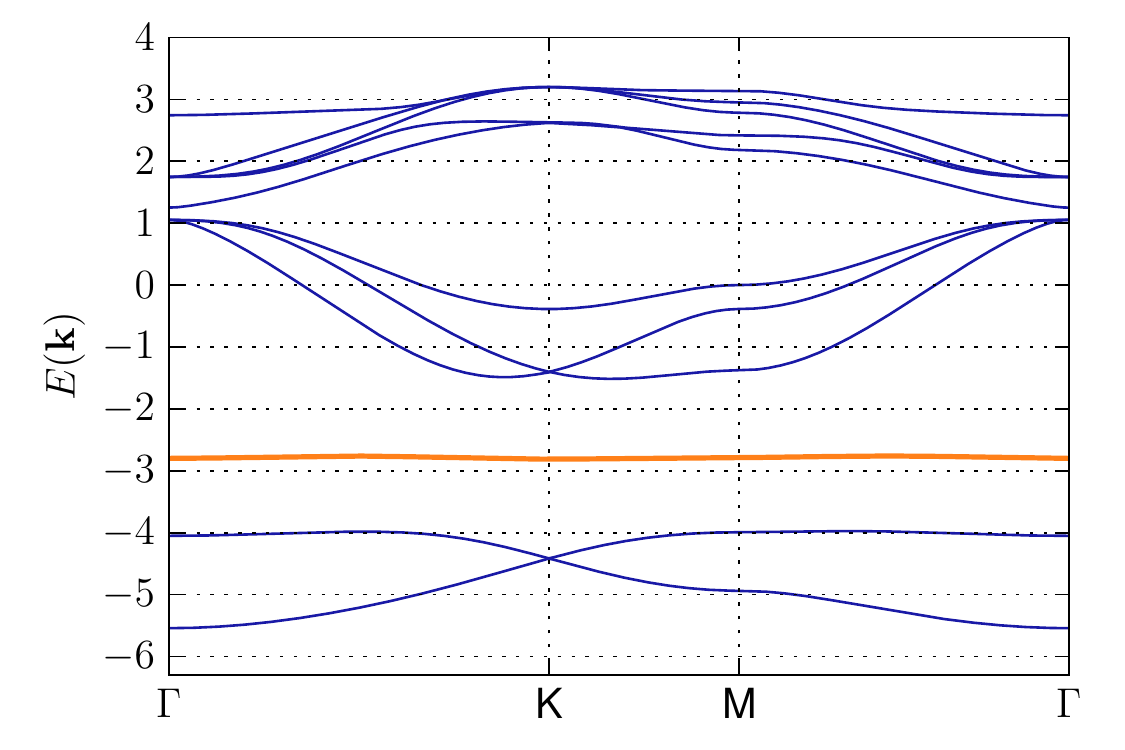}}
    \subfigure[~]{\includegraphics[width=0.33\textwidth]{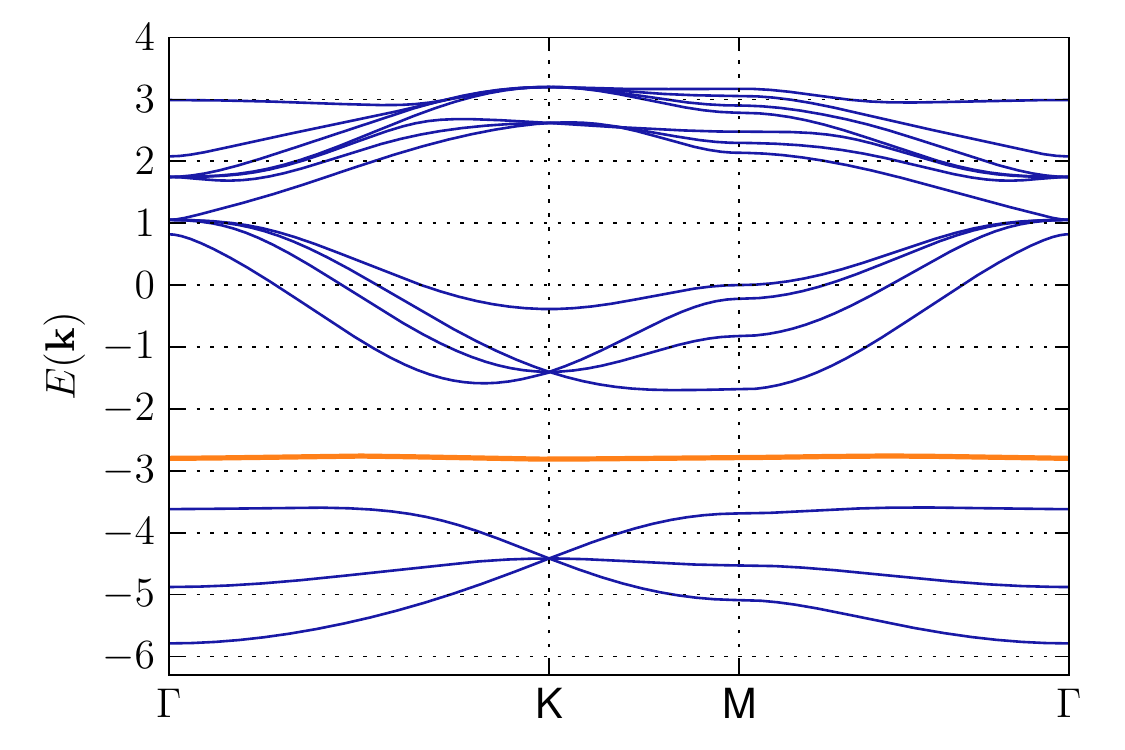}}
    \subfigure[~]{\includegraphics[width=0.33\textwidth]{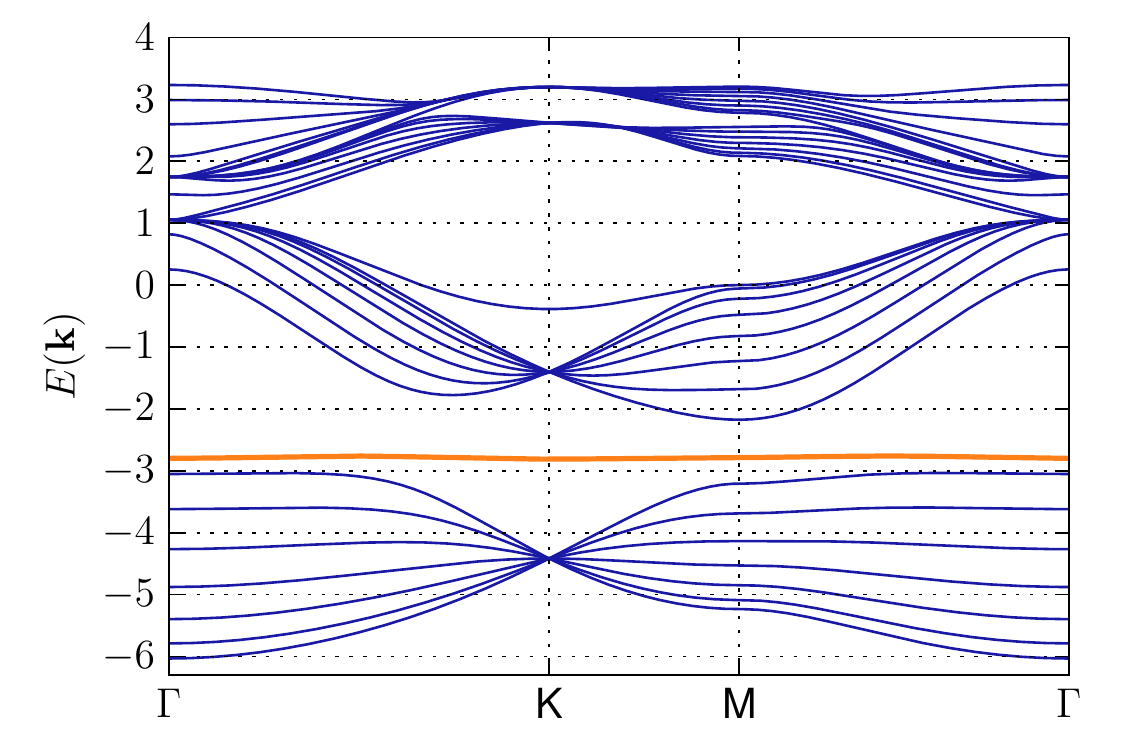}}
    \subfigure[~]{\includegraphics[width=0.33\textwidth]{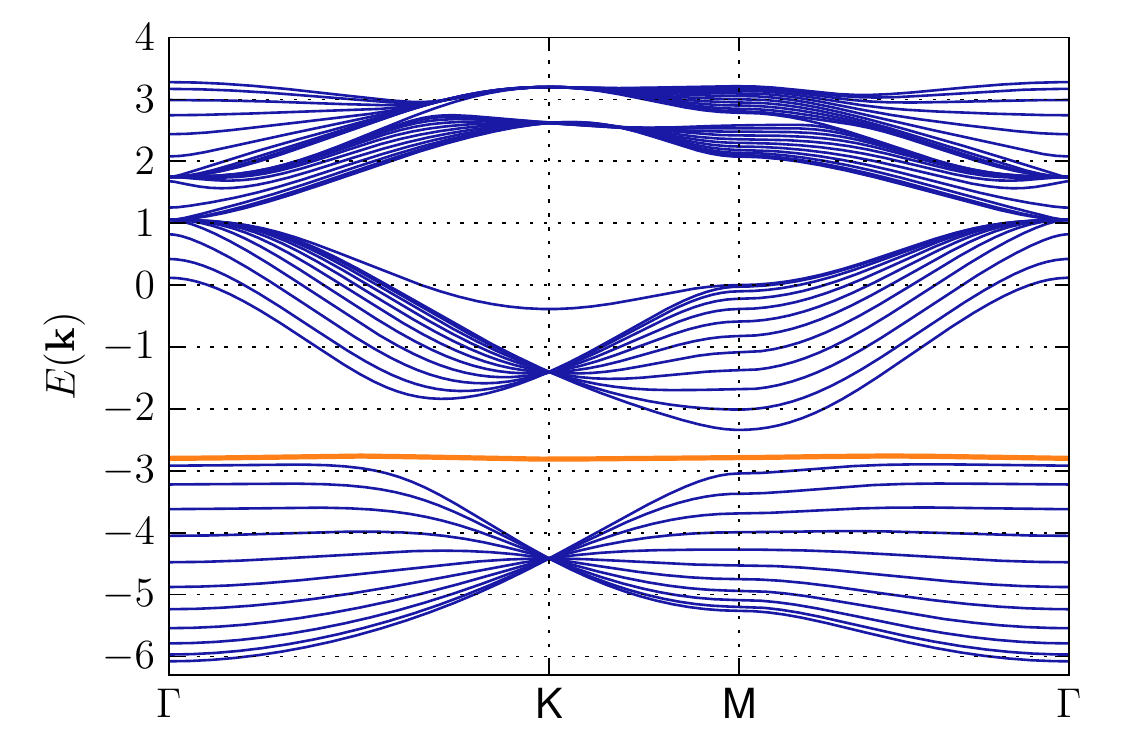}}
    \subfigure[~]{\includegraphics[width=0.33\textwidth]{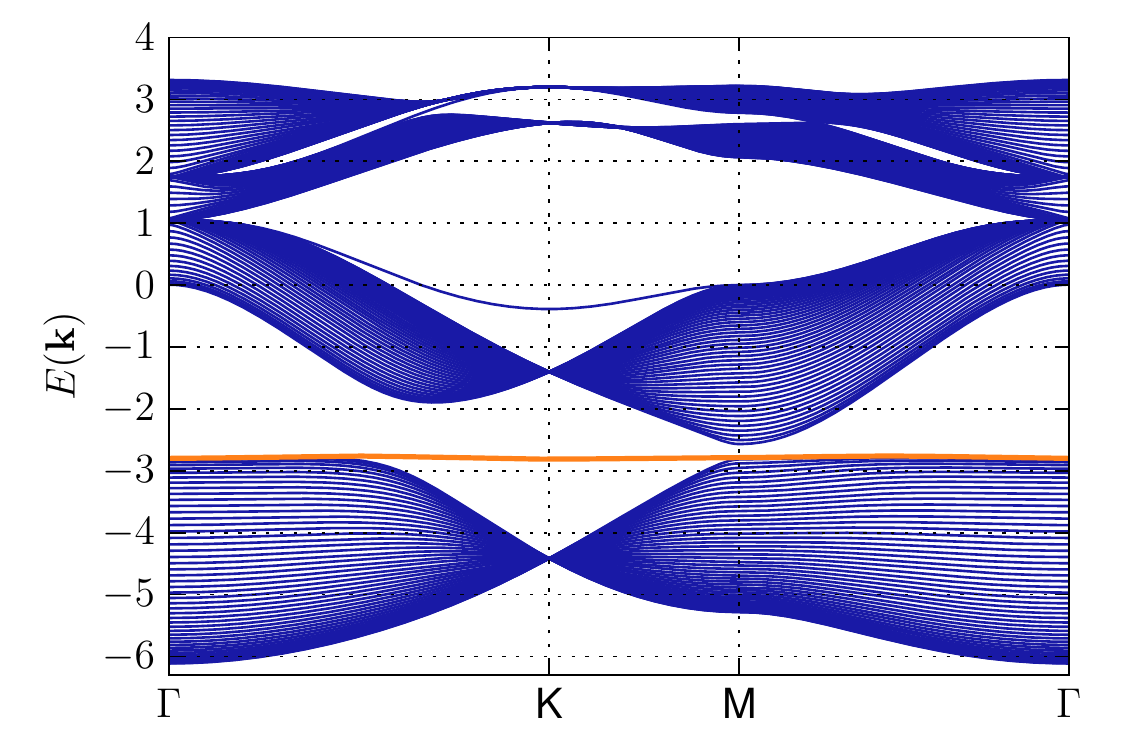}}
    \subfigure[~]{\includegraphics[width=0.33\textwidth]{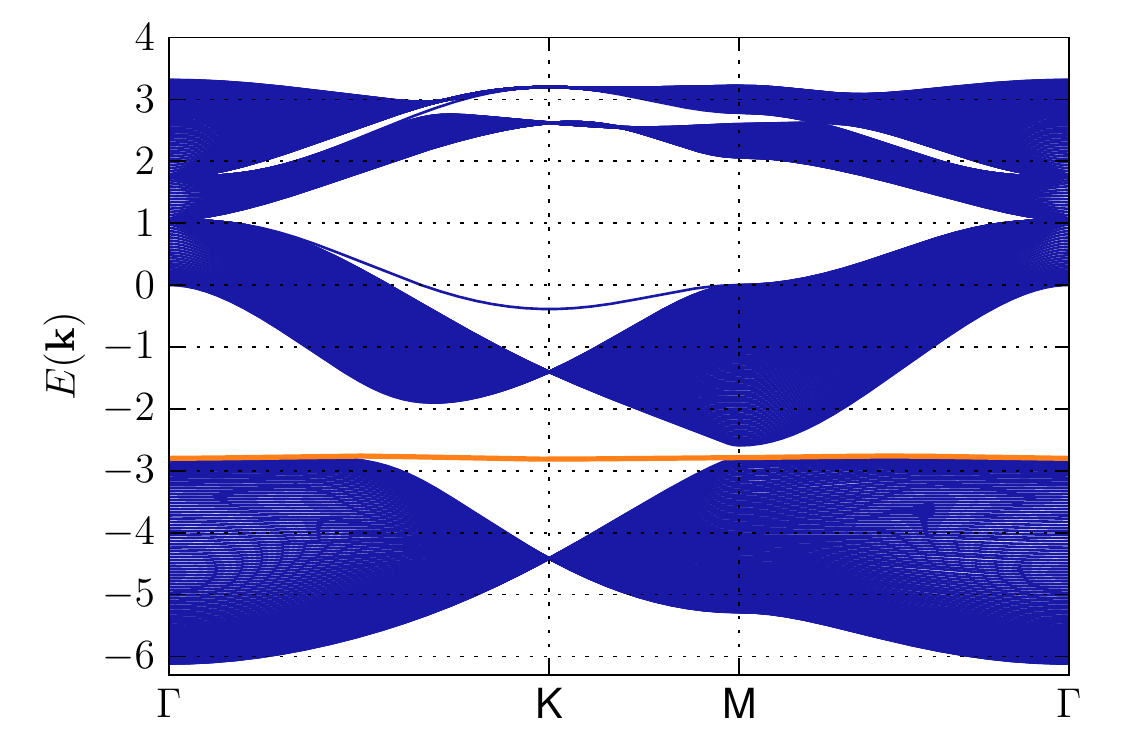}}
\caption{(Color online) Bulk dispersion, $E_s(\bb k)$, $s=1,2,\ldots,4N-1$, including topological flat bands in the presence of spin-orbit coupling. (a) shows a high-symmetry path through the Brillouin zone along which (b)-(i) show the dispersion of (\ref{ham}) for a system with $N=1,2,3,4,8,12,50,100$ stacked kagome layers respectively. In each case there is a very flat band (bold orange line) with Chern number $C=N$. The parameters are chosen as $t_1=-1,t_2=\lambda_1=0.3,\lambda_2=0.2, t_\perp=1.3$. As described in the text, the value of $t_\perp$ is not affecting the flat band but can be tuned to move the other bands (thin blue).} 
\label{fig:bands}
\end{figure*}

Here, we connect these research directions and consider a tight-binding model describing spin-orbit coupled electrons on a quasi-two-dimensional slab of pyrochlore including $N$ parallel kagome layers and $N-1$ intermediate triangular layers (Fig. \ref{fig:lattice}). Our main result is that this model accommodates virtually flat bands carrying Chern number $C=N$ that are well-separated from all other bands, even for relatively large $N\sim 10$ (cf. Fig. \ref{fig:bands}), and as such, provides an intriguing platform for yet unexplored phases of matter. 

{\it Setup.}
Our starting point is the following highly idealized model describing Rashba spin-orbit coupled electrons on pyrochlore slabs including $N$ kagome layers, $\mathcal K_m, m=1,\ldots,N$: 
  \begin{equation}
   H =\! \sum_{i,j,\sigma}t_{ij}c_{i\sigma}^{\dagger}c_{j\sigma}+\! i\! \sum_{i,j,\alpha,\beta} \! \lambda_{ij}(\bb{E}_{ij} \times \bb{R}_{ij}\! )\! \cdot\! \boldsymbol{\sigma}_{\alpha\beta} c_{i\alpha}^{\dagger}c_{j\beta} \ ,\label{ham}\end{equation}
 where $c_{i\sigma}^{\dagger}$ creates an electron with spin $\sigma$ at site $i$ and $\boldsymbol{\sigma}=(\sigma_1,\sigma_2,\sigma_3)$ is a vector containing the Pauli matrices as its elements. In the following we consider hopping, $t_{ij}$, between (next) nearest neighbors within each kagome plane with amplitude $t_1$ ($t_2$) as well as nearest-neighbor hopping, $t_\perp$, when involving the triangular layers [cf. Fig. \ref{fig:lattice}(a)]. Moreover, we consider only electrons whose spin is aligned in the $\hat{z}$ direction and consider the electric field $\bb{E}_{ij}$ originating from ions at the center of each hexagon in the kagome layer, $\mathcal K_m$, to effect only the hopping along the (directed) bonds, $\bb{R}_{ij}$, within $\mathcal K_m$, such that  $\lambda_{|i-j|}\equiv \lambda_{ij}|\bb{E}_{ij} \times \bb{R}_{ij}|$ parametrizes the nearest and next-nearest spin-orbit hopping strengths $\lambda_1$ and $\lambda_2$. For a single layer, $N=1$, our model (\ref{ham}) reduces to the kagome model introduced in Ref. \onlinecite{chernins1}.

In reciprocal space (\ref{ham}) is represented by a $(4N-1)\times (4N-1)$ Hermitian matrix, $\mathcal H_\bb k$, which is diagonalized, $\mathcal H_\bb k\ket{\psi_s(\mathbf k)}=E_s(\bb k)\ket{\psi_s(\mathbf k)}$, by the states $\ket{\psi_s(\mathbf k)}=\sum_ia^s_i(\bb k)\ket{i}$, where $\ket{i}$ labels the states in the unit cell, $s$ labels the bands in order of increasing energies, $E_{s+1}(\bb k)\geq E_{s}(\bb k)$, and $i,s=1,2,\ldots,4N-1$. For an explicit construction of $\mathcal H_\bb k$ we refer to the Supplemental Material. To characterize the bands with dispersion $E_s(\bb k)$ we calculate the Chern number, $C=\frac 1 {2\pi}\int_{\rm{BZ}} F^s_{12}({\mathbf k})d^2k$, which is a topologically protected integer quantity defined for an isolated band described by the wave functions $\ket{\psi_s(\mathbf k)}$, via the Berry curvature, $F^s_{ij}({\mathbf k})=\partial_{k_i}A^s_j({\mathbf k})-\partial_{k_j}A^s_i({\mathbf k})$, which in turn is defined in terms of the Berry connection $A^s_j({\mathbf k})=-i\bra{\psi_s(\mathbf k)}\partial_{k_j}\ket{\psi_s(\mathbf k)}$. The physical significance of the Chern number is that it counts the number of current carrying edge states, and as such gives the quantized Hall conductivity of a filled band, $\sigma_H=C\frac{e^2}{h}$ \cite{tknn} (and typically $\sigma_H=C\nu\frac{e^2}{h}$ for an incompressible state at fractional band filling, $\nu$).

{\it Band structure and wave functions.} 
We begin our discussion of the band structure of (\ref{ham}) by considering a few instructive limiting cases. 
First, for $t_1=\pm t_\perp, t_2=\lambda_1=\lambda_2=0$, there are $2N-1$ perfectly flat bands which can be understood from counting the number of linearly independent localized modes---for $N=1$ these can be taken as localized on each hexagon with amplitudes of alternating sign. (In the three-dimensional limit, $N\rightarrow \infty$, this is consistent with the two flat 3D bands known from bulk pyrochlore \cite{bergman}.)  These perfectly flat bands are, however, not carrying a well-defined Chern number as dispersive bands touch the flat ones at quadratic touching points occurring at $\bb{k}=0$. Next, for $t_\perp=0$ and finite spin-orbit coupling, $\lambda$, we get $N$ decoupled copies of the band structure, including $C=\pm 1$ bands, discovered in Ref. \onlinecite{chernins1} (in addition there are $N-1$  trivial bands due to the inert triangular layers).   

Turning to the full spin-orbit coupled model with $t_\perp\neq 0$, we quite generically find a band with Chern number $C=N$. Figure \ref{fig:bands} provides examples of the band structure, including flat bands with $C=N$ (in this example situated at $s=N$) for various $N$ given a suitable choice of hopping parameters. The $C=N$ bands have identical (weak) dispersion, $E_{s=N}(\bb k)$, regardless of the number of kagome layers, $N$, as will become clear later when we consider the properties of the wave functions. 
\begin{figure}
    \subfigure[~]{\includegraphics[width=\linewidth]{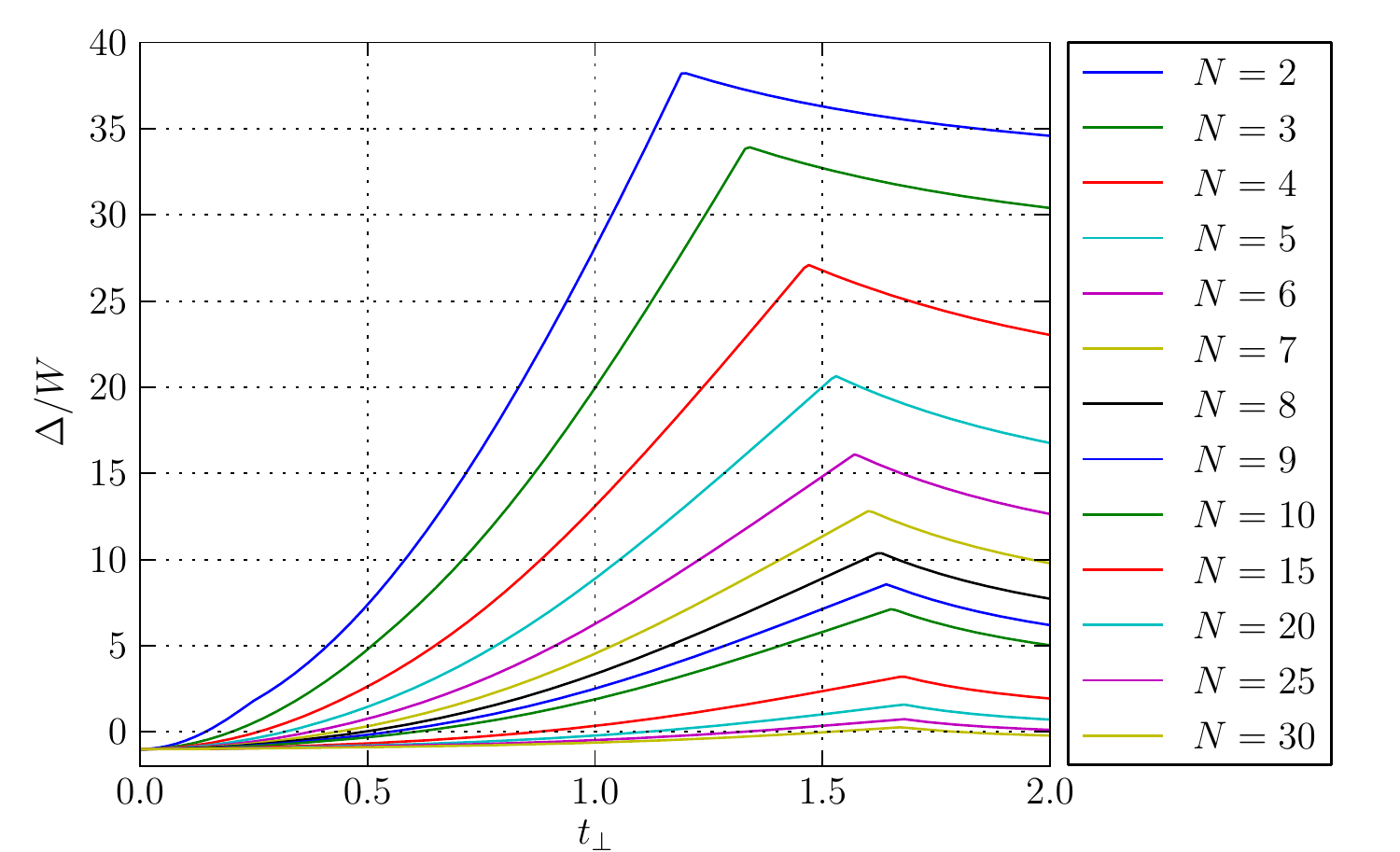}}
    \subfigure[~]{\includegraphics[width=\linewidth]{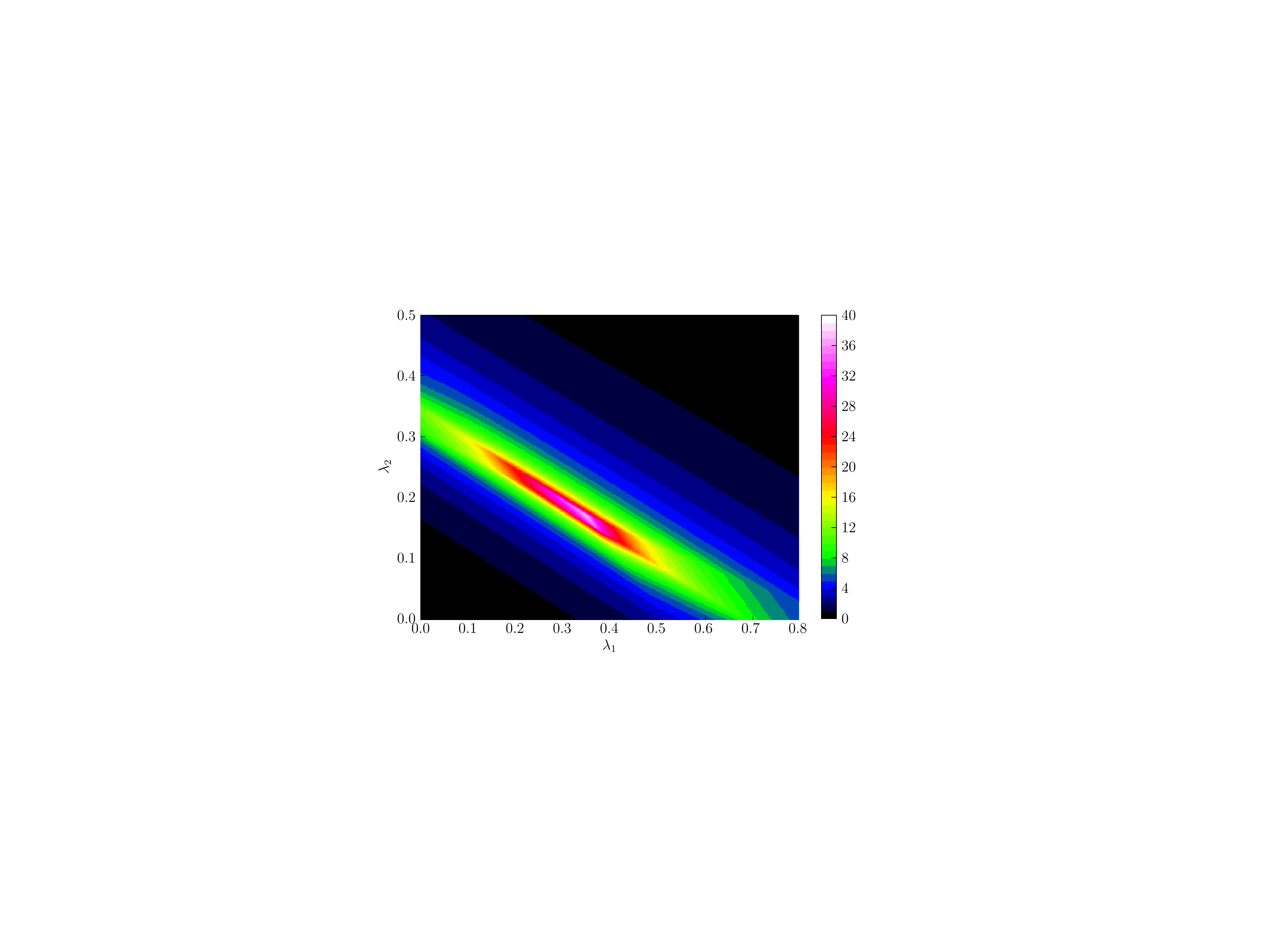}}
\caption{(Color online) The flatness ratio $\Delta/W$ (defined in the text) as a function of $t_\perp$ for various $N$ (a). Notably, the flatness reaches its maximum for relatively large $t_\perp$. Here we use $t_1=-1, t_2=0.3, \lambda_1=0.348, \lambda_2=0.172$. The kink reflects that the upper bands become closer to the topological band than the lower bands. In (b) we plot $\Delta/W$ for $N=2$ kagome layers as a function of  $\lambda_1$ and $\lambda_2$ for $t_\perp=1.3,t_1=-1, t_2=0.3$. The flatness ratio is large for a sizable region and, in fact, positive throughout the plotted range. } 
\label{fig:tperp}
\end{figure} 

To quantify how well the nontrivial bands are separated from the other bands we define the flatness ratio, $\Delta/W$, in terms of the bandwidth, $W=\max\limits_{\bb{k}, \bb{k}' \in \rm{BZ}}[E_{s}(\bb{k})-E_{s}(\bb{k}')]$, and the energy gap, $\Delta=\min\limits_{\bb{k}, \bb{k}' \in \rm{BZ}}[E_{s}(\bb{k})-E_{s-1}(\bb{k}'), E_{s+1}(\bb{k}')-E_{s}(\bb{k})]$. 
A negative flatness ratio does not imply the existence of touching points, but it signals that there is no chemical potential such that only the topological band is partly filled (at the noninteracting level). While the topological band itself is independent of the value of $t_\perp$ as long as it is nonzero, the other bands depend in general on the detailed value of $t_\perp$, leading to a nontrivial $t_\perp$-dependence of $\Delta/W$ as shown in Fig. \ref{fig:tperp}(a). In Fig. \ref{fig:tperp}(b) we fix $t_\perp=1.3,t_1=-1, t_2=0.3, N=2$ and plot $\Delta/W$ as a function of $\lambda_1$ and $\lambda_2$. This further illustrates the fact that although reaching optimal values is a matter of fine-tuning, there are sizable parameter regions for which $\Delta/W\gg1$. The separation of energy scales demonstrated in Figs. \ref{fig:bands},\ref{fig:tperp} implies that there is a region of parameter space up to relatively large $N\sim10$, such that $\Delta\gg U\gg W$, for some interaction strength, $U$. Note however, that one can in principle always make the bandwidth, $W$, arbitrarily small by including longer range hopping \cite{chernins3}, thus the limit on $N$ is not a fundamental one, but rather an estimate on what might be achieved for reasonable fine-tuning. Moreover, adding a chemical potential on the triangular layers, $H\rightarrow H+\mu\sum_{i\notin \mathcal K_{m}}n_i$, provides additional freedom of shifting the other bands while leaving the flat $C=N$ band completely unaltered.\cite{bandmoving}
 As in the $C=1$ bands studied earlier, this clearly provides a promising platform for strongly correlated states. In particular, the $C=N>1$ bands could lead to new phenomena with no analog in continuum systems (Landau levels). 
 
A prominent feature of the wave functions in the flat topological bands is that they are entirely localized to the kagome layers. 
In fact, we find that the wave functions in the $C=N$ bands can be completely understood in terms of the states $\ket{\phi^{C=1}(\mathbf k)}$ in the nearly flat band with Chern number $C=1$ in the single-layer kagome model.\cite{chernins1}
More specifically, the $C=N$ states can be written as $\ket{\psi_{s=N}(\mathbf k)}=\sum_{m=1}^{N}\alpha_m(\bb k)\ket{\phi_m^{C=1}(\mathbf k)}$, where $\ket{\phi_m^{C=1}(\mathbf k)}$ is the $C=1$ state localized to $\mathcal K_m$. The coefficients $\alpha_m(\bb k)$ are uniquely determined by local constraints similar to those in localized modes appearing in frustrated hopping models without spin-orbit interactions (cf., e.g., Ref. \onlinecite{bergman}): The relevant amplitudes in the kagome layers obey a sum rule leading to a vanishing total amplitude for hopping to the triangular layers. In terms of the wave function amplitudes, this can be stated as $a^{s=N}_{i\notin \mathcal K_{m}}(\bb k)=\sum_{j\in\{ \mathcal K_m\cup \mathcal K_{m+1}\}}e^{i\theta_\bb k^j}a^{s=N}_j(\bb k)=0, \forall \bb k,m$, where $\theta_\bb k^j$ depends on the unit cell and gauge conventions.\cite{uc} In terms of $\alpha_m(\bb k)$, this requirement leads to a nontrivial probability distribution, $P_m(\bb k)=|\alpha_m(\bb k)|^2\propto e^{\pm 2m/\delta(\bb k)}$, of being in kagome layer $\mathcal K_m$, as is illustrated in Fig.  \ref{fig:prob} and derived in the Supplemental Material.

\begin{figure*}
    \subfigure[~]{\includegraphics[width=0.39\linewidth]{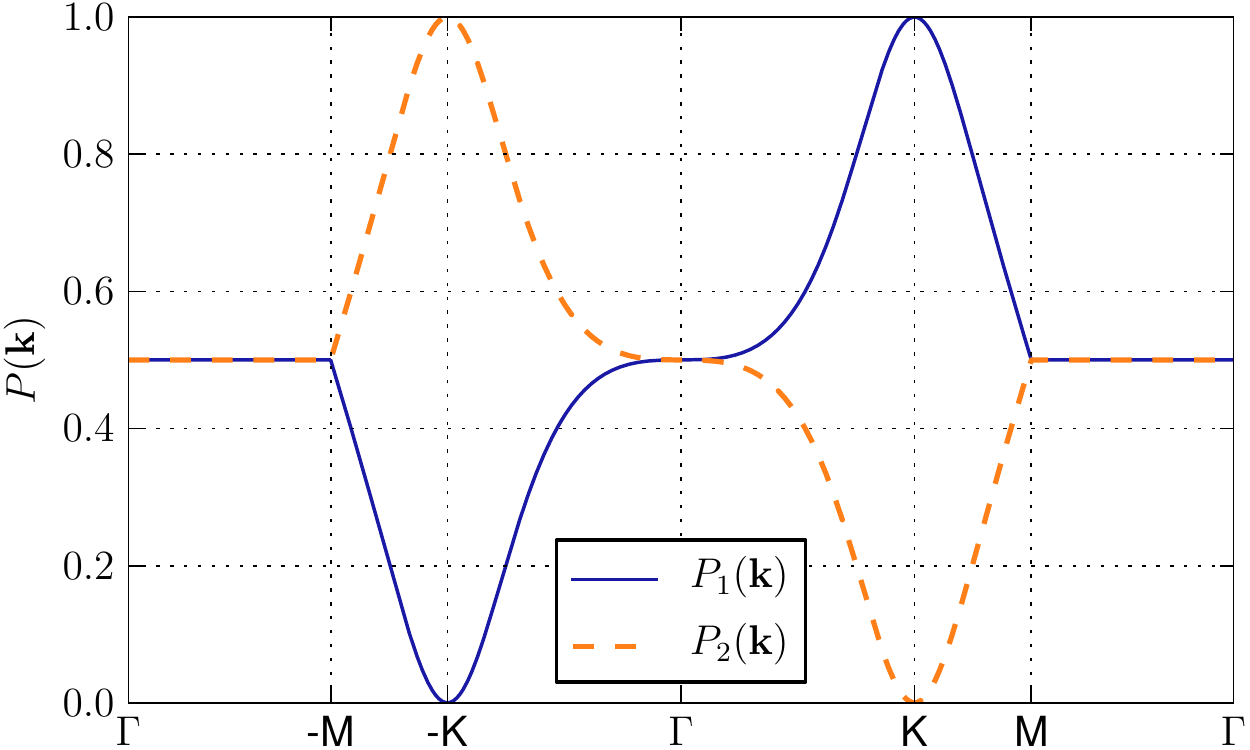}}
    \subfigure[~]{\includegraphics[width=0.39\linewidth]{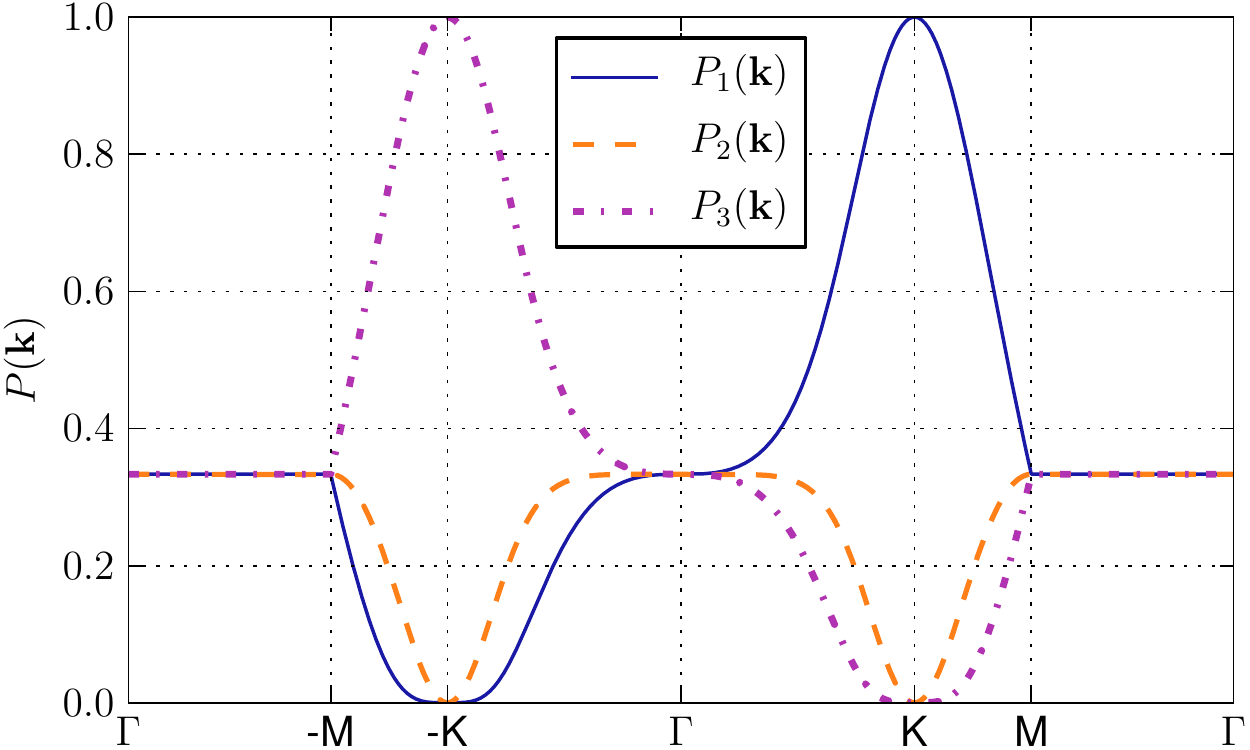}}
    \subfigure[~]{\includegraphics[width=0.39\linewidth]{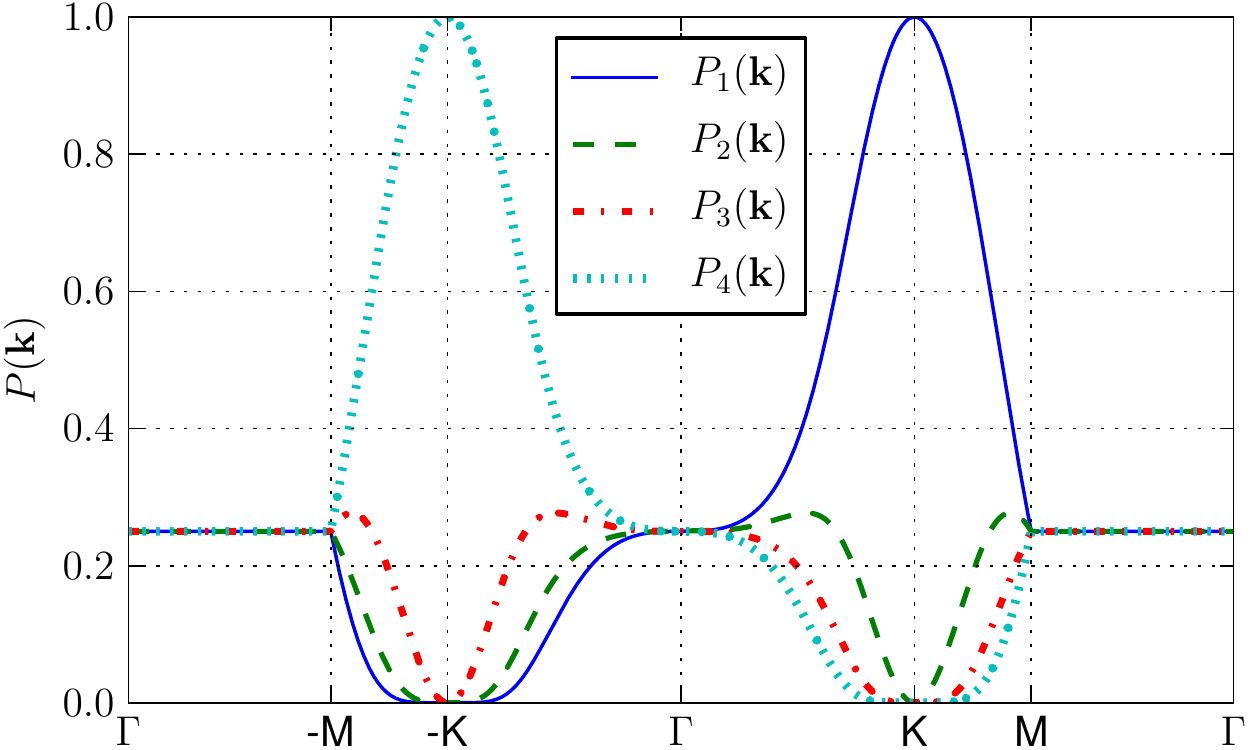}}
	\subfigure[~]{\includegraphics[width=0.39\linewidth]{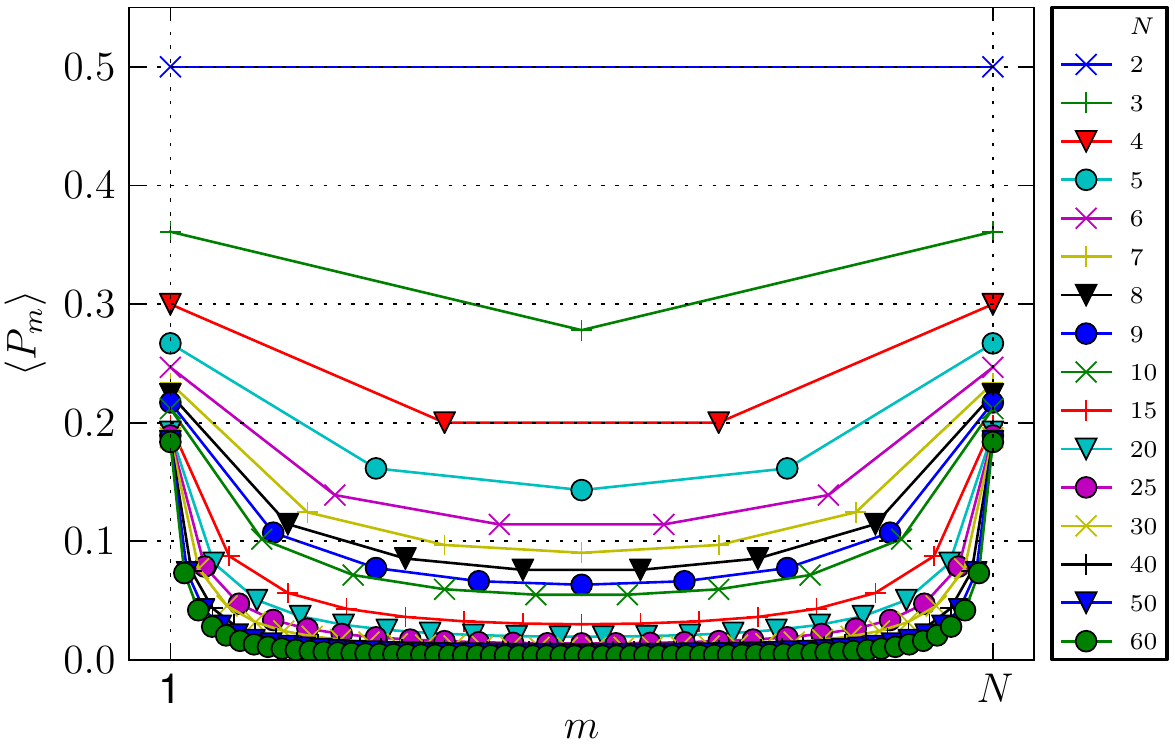}}
\caption{(Color online) The probability distribution, $P_m(\bb k)$, of being in different kagome layers, $\mathcal K_m$, in the $C=N$ topological bands ($t_\perp=1.3, t_1=-1, t_2=0.3, \lambda_1=0.3, \lambda_2=0.2$). In (a)-(c) we display the probability distributions for each layer within the $C=2,3,4$ bands respectively through a high symmetry path through the Brillouin zone (cf. Fig. \ref{fig:bands}(a)). That the wave functions are completely localized to the kagome layers is manifested by $\sum_mP_m(\bb k)=1, \forall \bb k$. The wave function is entirely localized to the top or bottom kagome layer at the $K$-points. In (d) we display the total weight, $\av{P_m}$, in each kagome layer for various $N$. Notably, most weight is carried by the outermost layers.} 
\label{fig:prob}
\end{figure*} 

That the wave functions are completely localized to the kagome layers is indeed reminiscent of the structure of multi-layer quantum Hall systems, or any system including $N$ decoupled $C=1$ bands for that matter. However, the structure of the Hilbert spaces is starkly different---while the interlayer dynamics in more conventional multilayer systems can in principle be independent, i.e., the layers are coupled only via interactions, the particle motion within different kagome layers in the $C=N$ bands is necessarily strongly coupled even in the absence of interactions between the layers. A manifestation of this coupling between layers is that translations in reciprocal space within the $C=N$ bands effectively amount to shifting the kagome layer. In Ref. \onlinecite{cn}, it was argued that bands with this type of symmetry enriched translation structure can, in principle, harbor entirely new phenomena such as wormholes with non-Abelian statistics.

The nontrivial translation structure is especially transparent in the $N=2$ case highlighted in Fig. \ref{fig:prob}(a): At the $K$-point the full weight of the wave function is in the lower layer, $P_1(\bb k=K)=1$, while it is entirely localized to the upper layer at $-K$, $P_2(\bb k=-K)=1$, i.e., at the $K'$-point. More generally, the role of the kagome layers $\mathcal K_m$ and $\mathcal K_{N+1-m}$ are interchanged by the transformation $\bb k \rightarrow -\bb k$. 

For $C=N>2$ the total weight is not evenly distributed over all kagome layers (Fig. \ref{fig:prob} (b)-(d)). For instance, as a function of $\bb k$, the probability of being in a certain layer reaches unity only in the top ($\mathcal K_N$) and bottom ($\mathcal K_1$) layers as a direct consequence of $P_m(\bb k)=|\alpha_m(\bb k)|^2\propto e^{\pm 2m/\delta(\bb k)}$. For large $N$ this implies that a considerable fraction of the integrated weight, $\av{P_m}=A^{-1}_{\rm{BZ}}\int_{\rm{BZ}}P_m(\bb k)d^2k$, is carried by the outermost layers, i.e., by the surface as is illustrated in Fig. \ref{fig:prob}(d). This surface localization is also in sharp contrast to more conventional multilayer systems.

{\it Discussion.}
In this work, we have demonstrated that nearly flat bands with Chern number $C=N$ exist in a short-range tight-binding model on a quasi-two-dimensional slab of the pyrochlore lattice. We have also provided a detailed characterization of the wave functions within these bands and highlighted their structure, which is qualitatively distinct from systems hosting multiple $C=1$ bands such as multilayer quantum Hall systems. At partial band filling it is very likely that this model harbors intriguing correlated states.  
Indeed, based on an elegant, albeit rather implicit, Wannier function approach \cite{qi}, it was recently argued\cite{cn} that 
lattice dislocations can lead to qualitatively new phenomena that are only realizable in $C>1$ bands. Here, we have provided a promising platform for realizing such phenomena and a testbed for future numerical studies of flat bands with variable Chern number. 

An intriguing open question is whether our model, {\it mutatis mutandis}, can be realized experimentally. In this context it is indeed promising that there is a rapid development in engineering quasi-two-dimensional (multi-layer) oxides\cite{oxideinterfaces}, and in particular, that there exists a number of pyrochlore-based transition-metal oxide materials, especially iridates, where the conduction electrons experience strong spin-orbit effects.\cite{pesin,weyl,fiete,kim} 

It has recently been shown that our model indeed harbors a large number of new fractional Chern insulator phases.\cite{ftihigher}

\acknowledgments

We acknowledge useful discussions with P. Brouwer, A. L\"auchli and Z. Liu. EJB also acknowledges related collaborations with A. L\"auchli, Z. Liu, and R. Moessner. EJB is supported by the Alexander von Humboldt foundation.

\newpage

\section*{Supplementary material for "Flat bands with higher Chern number in pyrochlore slabs"}
In this supplementary material, we provide details on how to construct the Hamiltonian in reciprocal space as well as a definition of the unit cell. 
We also derive an explicit expression for the $N$-layer $C = N$ states in terms of the single-layer $C = 1$ states and explain the exponential localization of the wave functions to the surface layers. Finally, we present Berry curvature data for various $C = N$ bands.

\subsection*{Unit cell, gauge conventions and construction of the Hamiltonian}

\begin{figure}[ht!]
    \begin{center}
        \includegraphics[width=\linewidth]{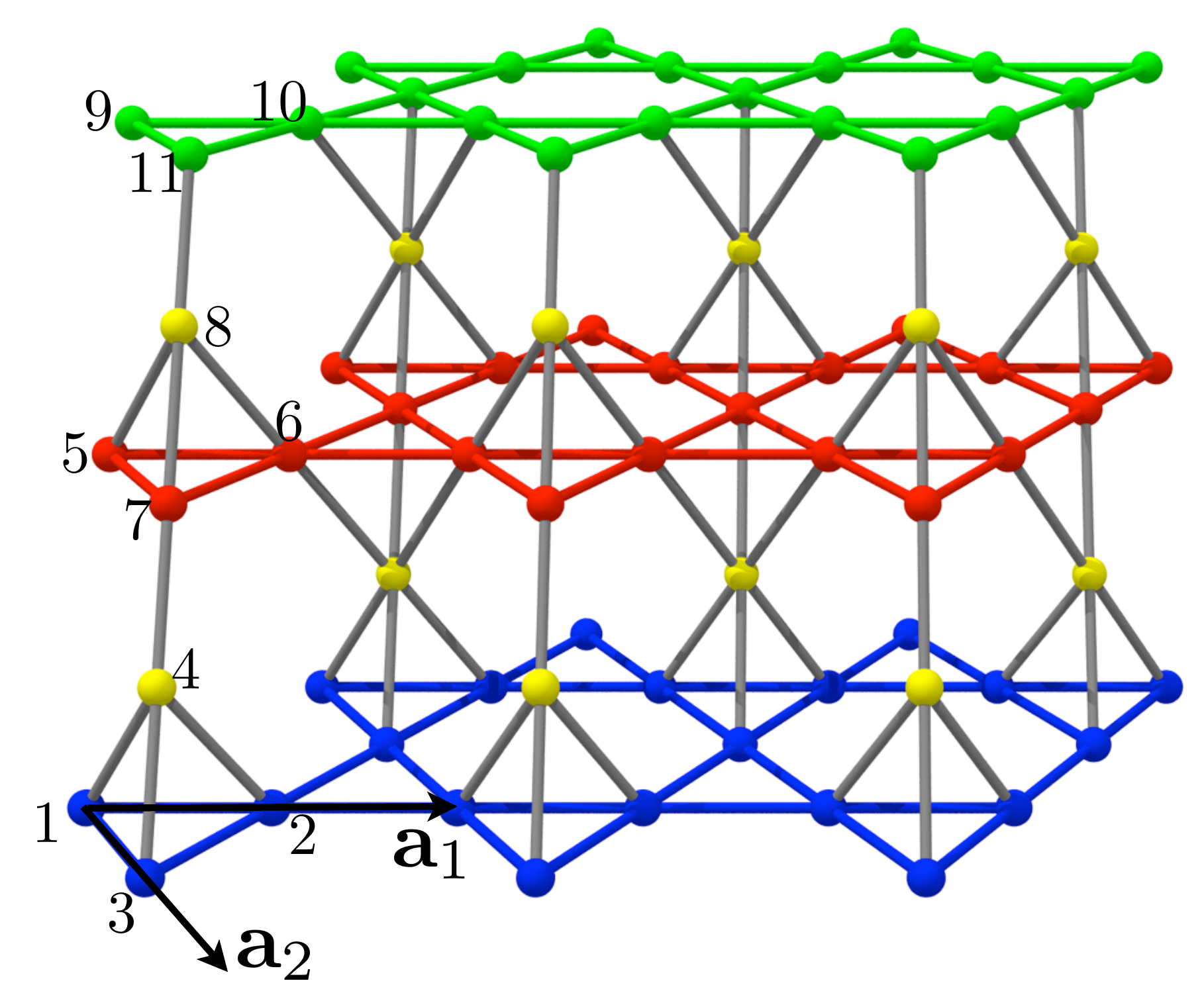}
    \end{center}
    \caption{Generalizable unit cell, shown for three layers. The basis vectors are $\bb a_1=2a\hat{x}, \bb a_2=a\hat{x}-\sqrt{3}a\hat{y}$, where $a$ is the lattice constant.}
    \label{fig:unitcell}
\end{figure}

To derive the reciprocal space Hamiltonian, $ \mathcal{H}_\bb k$ we need to fix the gauge and choose an appropriate unit cell. Here we chose a unit cell as depicted for $N=3$ in Fig. \ref{fig:unitcell} which has the benefit that the construction of $ \mathcal{H}_\bb k$ generalizes in a straight forward way to any number of kagome layers, $N$. Further, we use the gauge freedom such that hopping inside one unit cell does not give rise to a phase factor, while hopping a distance $\bb d=n\bb a_1+m\bb a_2$ (to another unit cell) gives rise to a phase factor $e^{i\bb k\cdot \bb d}$. With these conventions, and $k_i=\bb k \cdot \bb a_i, \ i=1,2$, $k_3=k_1-k_2$, the Hamiltonian for one kagome layer in reciprocal space reads
\begin{align}
    &\mathcal{H}^{N=1}_\bb k =  
    t_1
    \begin{pmatrix}
        0 & 1 + e^{i k_1} & 1 + e^{i k_2} \\
        1 + e^{- i k_1} & 0 & 1 + e^{-i k_3} \\
        1 + e^{- i k_2} & 1 + e^{i k_3} & 0 \\
    \end{pmatrix} \nonumber \\
    & + i \lambda_1 
    \begin{pmatrix}
        0 & 1 + e^{i k_1} & -(1 + e^{i k_2}) \\
        -(1 + e^{- i k_1}) & 0 & 1 + e^{-i k_3} \\
        1 + e^{- i k_2} & -(1 + e^{i k_3}) & 0 \\
    \end{pmatrix}
    \;  \nonumber \\
    \; &+  t_2
    \begin{pmatrix}
        0 & e^{i k_2} + e^{i k_3} & e^{i k_1} + e^{-i k_3} \\
        e^{- i k_2} + e^{-i k_3} & 0 & e^{-i k_1} + e^{i k_2} \\
        e^{- i k_1} + e^{i k_3} & e^{i k_1} + e^{-i k_2} & 0 \\
    \end{pmatrix} \nonumber \\
    & + i \lambda_2 
    \begin{pmatrix}
        0 & -(e^{i k_2} + e^{i k_3}) & e^{i k_1} + e^{-i k_3} \\
        e^{- i k_2} + e^{-i k_3} & 0 & -(e^{-i k_1} + e^{i k_2}) \\
        -(e^{- i k_1} + e^{i k_3}) & e^{i k_1} + e^{-i k_2} & 0 \\
    \end{pmatrix}, 
    \label{kh}
\end{align}
which is related to the form given in Ref. \onlinecite{chernins1} by a gauge transformation (and a change of basis). 

Our choice of unit cell makes the generalization to multiple coupled kagome layers especially simple as it allows us to reuse $\mathcal{H}^{N=1}_\bb k$ for each of the kagome layers. Moreover, the phases acquired when hopping to the triangular sites, $i=4,8,12,...,4N-4$, are the same for each triangular layer using these conventions.

For the two layer system one finds 
\begin{align}
    \mathcal{H}_\bb k^{N=2} =   
    \begin{pmatrix}
        & & & 1 & & & \\
        & \mathcal{H}^{N=1}_\bb k & & 1 &  & 0 & \\
        & & & 1 & & & \\ 
        1 & 1 & 1& 0 & e^{-i k_2} & e^{i k_3} & 1 \\
        & & & e^{i k_2} & & & \\
        & 0 & & e^{-i k_3} & &\mathcal{H}^{N=1}_\bb k & \\
        & & & 1 & & & 
    \end{pmatrix}\ ,
    \label{twolayer-H}
\end{align}
which can be more compactly by introducing
\begin{align}
    \mathcal{H}_{\perp} &= \begin{pmatrix} \mathcal{H}_{\perp,a} \\ 0 \\ \mathcal{H}_{\perp,b}\end{pmatrix}
                = t_\perp \begin{pmatrix} 1 \\ 1 \\ 1 \\ 0 \\ e^{i k_2} \\ e^{-i k_3} \\ 1\end{pmatrix} \ , 
    \label{mulilayer-connect}
\end{align}
with which we can write \eqref{twolayer-H} as:
\begin{align}
    \mathcal{H}^{N=2}_\bb k =   
    \begin{pmatrix}
        \mathcal{H}^{N=1}_\bb k & \mathcal{H}_{\perp,a} &  0\\
        \mathcal{H}_{\perp,a}^\dagger & 0  & \mathcal{H}_{\perp,b}^\dagger \\
        0 & \mathcal{H}_{\perp,b} & \mathcal{H}^{N=1}_\bb k  \\
    \end{pmatrix} \ .
    \label{twolayer-H-short}
\end{align}
This  generalizes directly to an arbitrary number of layers, $N$, yielding the following $(4N-1) \times (4N-1)$ matrix:

\begin{align}
    \mathcal{H}^{N}_\bb k =& \begin{pmatrix}
        \mathcal{H}^{N=1}_\bb k & \mathcal{H}_{\perp,a} &  0 & 0 & 0 & 0\\
        \mathcal{H}_{\perp,a}^\dagger & 0  & \mathcal{H}_{\perp,b}^\dagger & 0 & 0& 0\\
        0 & \mathcal{H}_{\perp,b} & \mathcal{H}^{N=1}_\bb k  & \dots & 0 & 0\\
        0 & 0 & \vdots &  \ddots & \mathcal{H}_{\perp,a} & 0 \\
        0 & 0 & 0 &  \mathcal{H}_{\perp,a}^\dagger & 0 & \mathcal{H}_{\perp,b}^\dagger \\
        0 & 0 & 0 & 0 &\mathcal{H}_{\perp,b} & \mathcal{H}^{N=1}_\bb k \\
    \end{pmatrix} \ .
    \label{multilayer-hamilton-visual}
\end{align}
The eigenstates in band $s$ are \begin{align}\psi_s(\mathbf k)=\begin{pmatrix} a^s_1(\bb k) \\ a^s_2(\bb k) \\ \vdots \\ a^s_{4N-1}(\bb k) \end{pmatrix}\ , \end{align}
which can also be written as $\ket{\psi_s(\mathbf k)}=\sum_ia^s_i(\bb k)\ket{i}$, as in the main text.

\subsection*{Wavefunction structure and surface localization}

With the conventions above, the constraints for the $C=N$ bands,
\begin{align}
\sum_{j\in\{ \mathcal K_m\cup \mathcal K_{m+1}\}}{e^{i\theta_\bb k^j}a^{s=N}_j(\bb k)} &= 0 & \forall \bb k,m , 
    \label{constraint}
\end{align}
where $\mathcal{K}_m\cup \mathcal{K}_{m+1} = \{ 3m-2, 3m-1, 3m, 3m+2, 3m+3, 3m+4 \}$
are fulfilled for
\begin{align}
    \theta_\bb k^j &= -k_2 & \textrm{for } j = 3m+2 \\ 
    \theta_\bb k^j &= k_3  & \textrm{for } j = 3m+3 \\
    \theta_\bb k^j &= 0 & \textrm{otherwise.}
    \label{}
\end{align}
We can now explicitly express the state $\ket{\psi_{s=N}(\bb{k})}=\sum_{m=1}^{N}\alpha_m(\bb k)\ket{\phi_m^{C=1}(\mathbf k)}$ 
in terms of the single layer eigenstate $\ket{\phi_m^{C=1}(\mathbf k)} := (a_1(\bb{k}) , a_2(\bb{k}), a_3(\bb{k}))^T$.
Together with \eqref{constraint} this leads to the relation
\begin{align}
    \frac{\alpha_{m+1}(\bb{k})}{\alpha_m(\bb{k})} &= \frac{-\left(a_1(\bb{k}) + a_2(\bb{k}) + a_3(\bb{k}) \right)}{e^{-i k_2}a_1(\bb{k}) + e^{i k_3}a_2(\bb{k}) + a_3(\bb{k})}
    \equiv r(\bb{k}) \, .
    \label{alpharatio}
\end{align}
We emphasize that the ratio $r(\bb{k}) = \frac{\alpha_{m+1}(\bb{k})}{\alpha_m(\bb{k})}$ does not depend on $m$.
From this we directly get 
\begin{align}
    \alpha_m(\bb{k}) &= \alpha_0(\bb{k}) r(\bb{k})^m  \, ,
\end{align}
where $\alpha_0(\bb{k})$ is set by normalization. One finds
\begin{align}
    1 &= \sum_{m=1}^N{|\alpha_m(\bb{k})|^2} = \sum_{m=1}^N{|\alpha_0(\bb{k}) r(\bb{k})^m|^2} \nonumber \\
    &= |\alpha_0(\bb{k})|^2 \sum_{m=0}^{N-1}{(|r(\bb{k})|^2)^{m+1}} \\
    \Rightarrow \quad |\alpha_0(\bb{k})|^2 &= \frac{1}{|r(\bb{k})|^2} \frac{|r(\bb{k})|^2-1}{(|r(\bb{k})|^2)^N - 1} \, ,
    \label{a_0}
\end{align}
except for $|r(\bb{k})|^2 = 1$ where $\alpha_0 = 1/N$. Now we can write the eigenstates in the $C=N$ band as
\begin{align}
    \ket{\psi_{s=N}(\mathbf k)}=\sum_{m=1}^{N}\alpha_0(\bb k) r(\bb{k})^m \ket{\phi_m^{C=1}(\mathbf k)} \, .
\end{align}
i.e., as an entangled sum of the wave functions in the single-layer $C=1$ bands.

To highlight the surface localization we define the $\bb k$-dependent localization length, $\delta(\bb{k}) = 1/|\log(|r(\bb{k})|)|$,
which leads to 
\begin{align}
    P_m(\bb{k}) = |\alpha_m(\bb{k})|^2 = |\alpha_0(\bb{k})|^2 \, e^{\pm 2m/\delta(\bb{k})} \, ,
    \label{exponentiall-probability}
\end{align}
as stated in the main text.

\subsection*{Berry curvature}

\begin{figure*}
    \subfigure[~]{\includegraphics[width=0.49\linewidth]{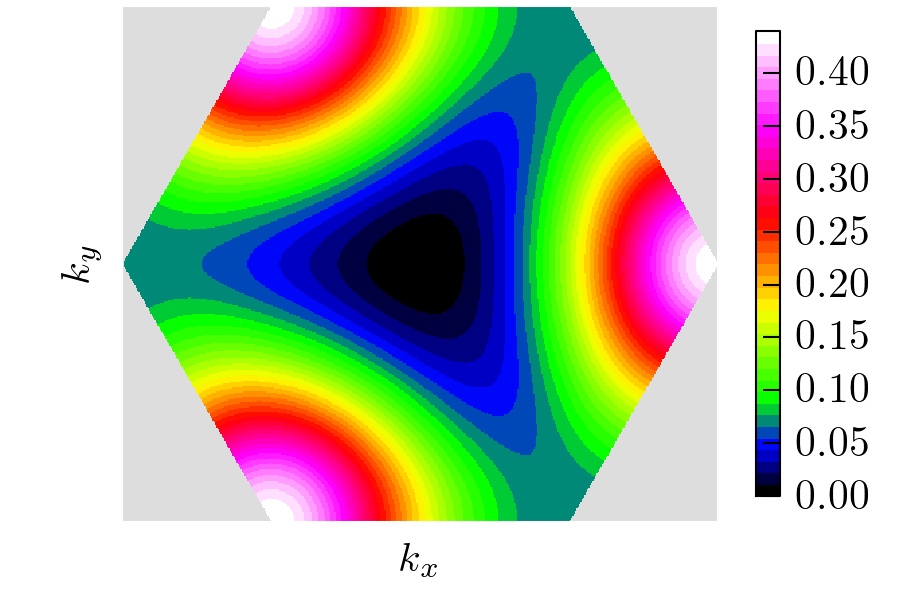}}
    \subfigure[~]{\includegraphics[width=0.49\linewidth]{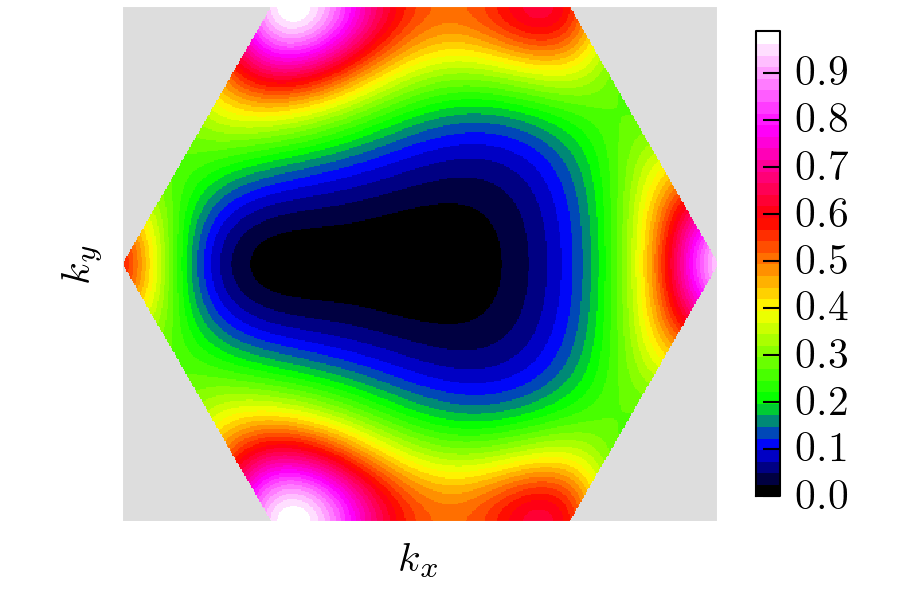}}
        \subfigure[~]{\includegraphics[width=0.49\linewidth]{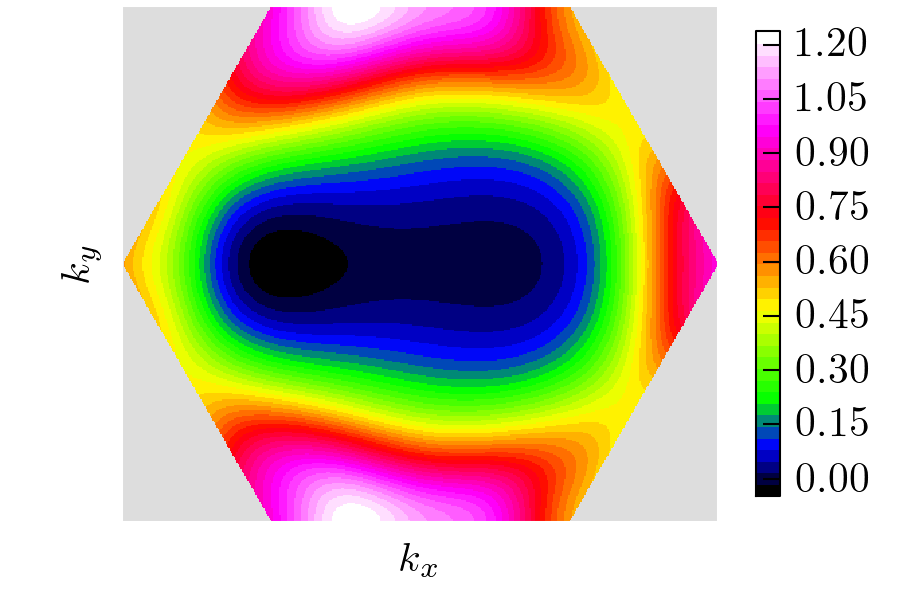}}
    \subfigure[~]{\includegraphics[width=0.49\linewidth]{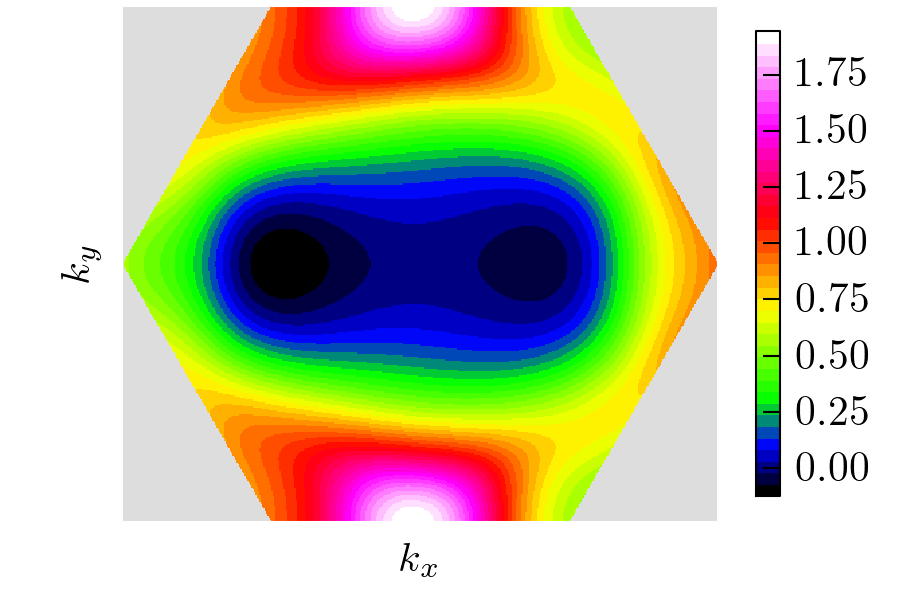}}
    \subfigure[~]{\includegraphics[width=0.49\linewidth]{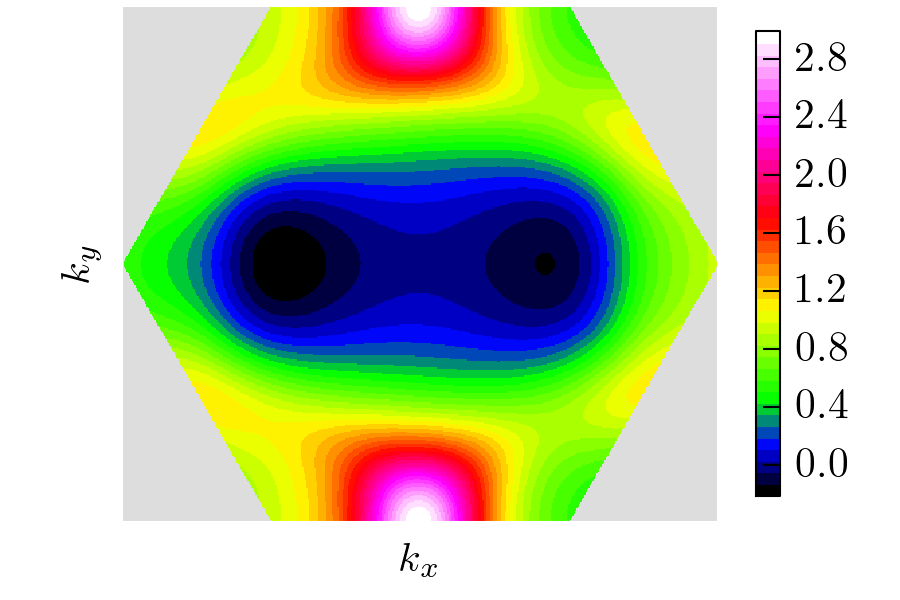}}
	\subfigure[~]{\includegraphics[width=0.49\linewidth]{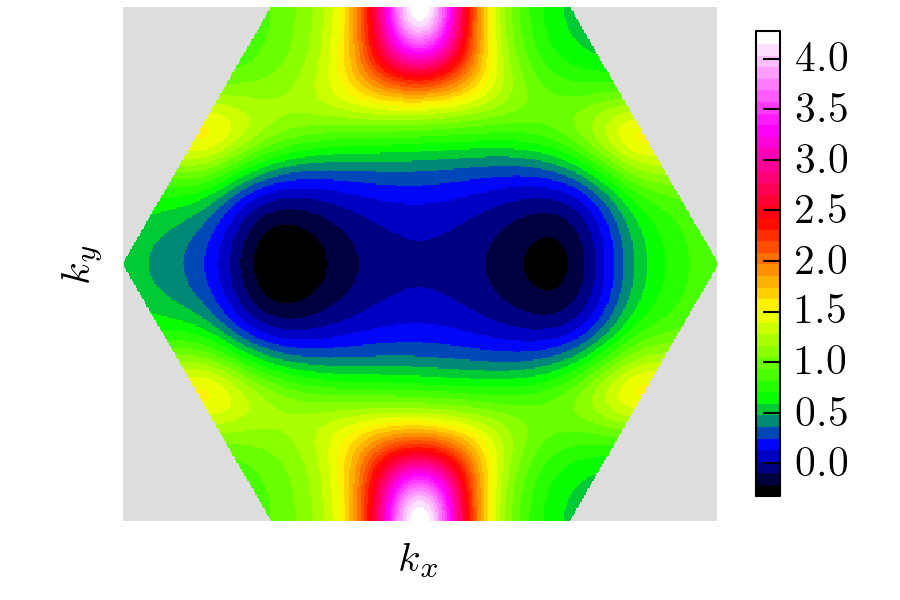}}

\caption{(Color online) Berry curvature in the $C=N$ bands with $t_1=-1, t_2=0.3, \lambda_1=0.3, \lambda_2=0.2$ (independent of $t_\perp$). In (a)-(f) we display the results for $N=1,2,3,...,6$ respectively. } 
\label{fig:berry}
\end{figure*}

In Fig. \ref{fig:berry}, we display the Berry curvature for various $N$ given the conventions specified above. 
Compared to the $C=N=1$ band (Fig. \ref{fig:berry}(a)) the maximum of the Berry curvature gradually shifts position for increasing $N$ (Fig. \ref{fig:berry}(b)-(f)) . Meanwhile, it remains smooth, positive (semi-)definite and free from singularities. 

%
%


\begin{thebibliography}{99}
\bibitem{multilayers} Topologically, a band with Chern number $N$ is equivalent to $N$ Landau levels as both systems have $N$ chiral edge states, but there are important differences as discussed here and in Ref. \onlinecite{cn}.
\bibitem{bandmoving} Generically $\mu>0$ moves other bands upwards (and changes the eigenstates) while the $C=N$ bands are completely unaffected as the states in these bands have a vanishing amplitude on the triangular lattice sites.
\end{thebibliography}
\end{document}